\documentclass[12pt,preprint]{aastex}

\usepackage{graphicx}
\usepackage{txfonts}
\def\ion#1#2{{\rm #1}\,{\sc #2}}

\newcommand{\etal}{et al.}
\slugcomment{Submitted to the Astrophysical Journal Letters}
\shorttitle{Temporal variation in excited Fe$^+$ near a GRB afterglow}
\shortauthors{Dessauges-Zavadsky \etal.}
\begin{document}
\title{TEMPORAL VARIATION IN THE ABUNDANCE OF EXCITED Fe$^+$ NEAR A GRB 
AFTERGLOW}

\author{
  Miroslava Dessauges-Zavadsky\altaffilmark{1},
  Hsiao-Wen Chen\altaffilmark{2},
  Jason X. Prochaska\altaffilmark{3},
  Joshua S. Bloom\altaffilmark{4},
  Aaron J. Barth\altaffilmark{5}
}

 \altaffiltext{1}{Observatoire de Gen\`eve, 51 Ch. des Maillettes, 
	1290 Sauverny, Switzerland;
	miroslava.dessauges@obs.unige.ch}
\altaffiltext{2}{Department of Astronomy \& Astrophysics; University of 
	Chicago; 5640 S.\ Ellis Ave., Chicago, IL 60637; 
	hchen@oddjob.uchicago.edu}
\altaffiltext{3}{UCO/Lick Observatory; University of California, Santa Cruz;
	Santa Cruz, CA 95064; xavier@ucolick.org}
\altaffiltext{4}{Department of Astronomy, 601 Campbell Hall, 
        University of California, Berkeley, CA 94720-3411; 
	jbloom@astron.berkeley.edu}
\altaffiltext{5}{Department of Physics and Astronomy, University of California 
	at Irvine, 4129 Frederick Reines Hall, Irvine, CA 92697-4575;
	barth@uci.edu}

\begin{abstract}

Excited Si$^+$ and Fe$^+$ species are routinely observed in the host
environment of $\gamma$-ray burst (GRB) afterglows, but are not
commonly seen in other extragalactic locations.  Their presence
signals unusual properties in the gaseous environment of these GRB
hosts that arise either as a result of the intense ionizing radiation
of the afterglow or through collision excitation in a dense cloud.  In
particular, the photon pumping scenario has explicit expectations for
temporal variation in the strength of the excited lines, owing to the
decline in the ionizing flux of the GRB afterglow.  We analyze
afterglow spectra of GRB\,020813 obtained in two epochs of
$\approx$\,16 hours apart, and
examine transitions from the first excited state of Fe$^+$ at $J=7/2$
in these two sets of data. We report a significant decline by at least
a factor of five in the equivalent width of the
\ion{Fe}{ii}\,$\lambda$2396 transition, the strongest from the $J=7/2$
state. We perform a Monte-Carlo analysis and determine that this
temporal variation is present at more than 3\,$\sigma$ level of
significance. This observation represents the first detection in the 
temporal variation of the excited Fe$^+$ states in the GRB host ISM, a 
direct influence of the burst itself on its environment.  We further
estimate that the Fe$^+$ gas resides in 50-100 pc from the afterglow,
based on the afterglow lightcurve and the presence and absence of the
excited \ion{Fe}{ii}\,$\lambda$2396 in the two-epoch observations.

\end{abstract}


%

\section{Introduction}

Long-duration $\gamma$-Ray Bursts (GRBs) are believed to trace the
death of massive, short lived stars \citep[see][ for review]{woosley06}. 
The brightness of GRB afterglows for a brief period makes them visible 
throughout most of the observable Universe \citep[e.g.][]{kawai06} 
and therefore a sensitive probe of both the interstellar medium (ISM) 
of their host galaxies and the intervening gas 
\citep[e.g][]{vreeswijk04,chen05,prochter06}. The study of metal 
absorption lines detected in the afterglow spectra illuminates the 
physical conditions in the ISM hosting the GRB, such as the \ion{H}{i} 
column density, the gas metallicity, the dust-to-gas ratio, and the 
kinematics.  Moreover, the access to features belonging to the 
circumstellar medium of the progenitor star of the GRB may yield strong 
constraints on the progenitor and may help to understand the impact of 
the GRB on the local environment (e.g.\ Ramirez-Ruiz \etal\ 2005; van 
Marle, Langer, \& Grac\'ia-Segura 2005; Prochaska, Chen, \& Bloom 2006, 
hereafter PCB06).

Interestingly, the transitions from excited states of C$^+$, Si$^+$,
and O$^0$ appear to be a generic feature in GRB host environments,
while they are observed through only rare sightlines, such as broad
absorption-line (BAL) quasars \citep{hall02}, and in gas near stars
like $\eta$~Carinae and $\beta$~Pictoris \citep{gull05,lagrange88}. In
addition, a large fraction of the GRB hosts also show excited Fe$^+$
lines \citep{prochaska06}. Does the detection of these excited state
absorption lines reveal extreme (pre) circumburst environments or are
they the result of work done by the afterglow in early times after the
burst?  To answer this we need to consider competing scenarios for the
production of excited-state transitions, and particularly for Fe$^+$:
(1)~collisional excitation and (2)~photon pumping. PCB06 have recently
shown that in the presence of an intensified UV radiation field from
the afterglow --~known to be stronger than $10^5$ to $10^6$ times the
ambient far-UV radiation field in the ISM of the Milky Way~-- indirect
UV pumping alone can produce the excited states observed in GRB hosts
without imposing extreme gas density and temperature\footnote{A third 
scenario is for the bright progenitor star, such as a WR star, to UV 
pumping these excited Fe$^+$. However, as we will show below, at the 
distance allowed for Fe$^+$ to survive the intense afterglow UV 
radiation ($\sim 50$ pc) the excitation rate from the WR star is 
insufficient to produce the observed excited states. In addition, the 
excitation rate due to direct IR pumping by the afterglow is orders of 
magnitude smaller than the excitation rate due to afterglow UV pumping.}.  
Under the photon pumping conditions, temporal variation in the 
absorption-line strength of excited state transitions is expected, 
because of the decline in the ionizing flux of the afterglow. 
\citet{perna98} and \citet{bottcher99} predicted time-dependent 
resonance absorption features in the GRB afterglow spectra, such as the 
\ion{Mg}{ii} doublet, over a timescale of a few days to a few weeks. 
Their analysis also indicated that important constraints on the extent 
and gas density of the circumburst medium can be derived. However, 
searches for absorption-line variability have so far yielded null 
results \citep{vreeswijk01,mirabal02}.

Under the photon pumping scenario, temporal variation in excited
transitions on scales of a few hours is expected, because the decay
timescale for the excited states ranges from 8 to 15 minutes. Given 
the new insights into the origin of excited lines, in this {\em Letter} 
we analyze afterglow spectra of GRB\,020813 obtained in two different 
epochs of $\approx\,16$ hours apart.  These data sets have a suitable 
time coverage to serve the purpose to search for variation in the 
excited Fe$^+$ lines.  We report a significant decline in the equivalent 
width of the strongest transition from the excited state Fe$^+$ $J=7/2$.  
This observation represents the first detection in temporal variation of 
the excited Fe\,II states in GRB host ISM, lending strong support for 
the indirect UV pumping scenario as the primary excitation mechanism 
of this gas.
%

\section{Observations and Data Reduction}

GRB\,020813 was detected by the High Energy Transient Explorer 2
(HETE\,2) at 2:44 UT on 2002 August 13 \citep{villasenor02,fox02}. Two
sets of spectroscopic observations were obtained for the optical
afterglow: spectropolarimetery with the Low Resolution Imaging
Spectrograph \citep[LRIS;][]{oke95,goodrich95} on the Keck~I telescope
and echelle spectroscopy with the Ultraviolet-Visual Echelle
Spectrograph \citep[UVES;][]{dodorico00} on the VLT Kueyen ESO
telescope. The epoch~1 LRIS data were acquired on 2002 August 13,
starting at 7:23 UT \citep{barth03} and spanning 3 hours in
duration. The brightness of the GRB at 4.7 hours after the burst was
$R\approx 18.4$. The observed spectral range is $\lambda = 3200-9400$
\AA\ with a wavelength scale of 1.09 \AA~pixel$^{-1}$ in the blue half
and 1.86 \AA~pixel$^{-1}$ in the red half. The signal-to-noise ratio
per pixel achieved by combining the twelve 15 minute exposures is very
good, S/N $= 50-100$.

The epoch~2 UVES data were acquired on 2002 August 13, starting at
23:32 UT 16.1 hours after the start of the LRIS observations and
spanning roughly 2.1 hours in duration \citep{fiore05}. The brightness
of the GRB was then $R\approx 20.4$. Two exposure times of 2300 and
3600~s were obtained with the dichroic \#1 B346+R580 and one exposure
of 1800~s with the dichroic \#2 B437+R860. We reduced these spectra,
available in the ESO Science Archive (program ID 69.A-0516), using the
ESO data reduction package {\tt MIDAS} and the UVES pipeline in an
interactive mode. A detailed description of the pipeline can be found
in \citet{ballester00}. To optimize the results, we made a systematic
check of each step of the pipeline reduction. Individual exposure
spectra were weighted by their S/N ratios and co-added to form a final
stacked spectrum. The resulting UVES high-resolution spectrum extends
from 4800 to 6800 \AA\ and has a wavelength scale of 0.029
\AA~pixel$^{-1}$ in the blue half and 0.035 \AA~pixel$^{-1}$ in the
red half. The achieved signal-to-noise ratio per pixel is low, S/N $=
4-5$ (the S/N is even poorer at $4800 > \lambda > 6800$ \AA).
%

\section{Temporal Variation}

\citet{barth03} identified two absorption systems on the GRB\,020813
line of sight at $z = 1.255$ and $z = 1.224$ with the former being
associated with the host galaxy of the GRB. \citet{savaglio04}
performed a detailed abundance analysis of the GRB\,020813 host
ISM. They obtained the column density measurements of numerous ions
detected in the LRIS spectrum --~\ion{C}{i}, \ion{Mg}{ii},
\ion{Mg}{i}, \ion{Al}{ii}, \ion{Si}{ii}, \ion{Ti}{ii}, \ion{Cr}{ii},
\ion{Mn}{ii}, \ion{Fe}{ii}, \ion{Ni}{ii}, \ion{Zn}{ii}, and
\ion{Ca}{ii}. The high-resolution UVES spectrum was analyzed by
\citet{fiore05}, where only the \ion{Mg}{ii}, \ion{Mg}{i}, and
\ion{Fe}{ii} ions are accessible.
%

%

An emerging feature of GRB environments is the presence of excited
state transitions \citep{vreeswijk04,chen05,berger06,prochaska06}. For
the host of GRB\,020813, only the detection of the excited
\ion{Si}{ii}\,$\lambda$1533 ($J=3/2$) transition was reported so far
\citep{savaglio04}.  We show, in addition, that the excited
\ion{Fe}{ii}\,$\lambda$2396 ($J=7/2$) transition, the strongest of all
from the $J=7/2$ state (i.e., with the highest $\lambda f$ product),
is detected.  We report in Table~\ref{EW} measurements and upper
limits for the rest-frame equivalent widths of various transitions
from the $J=7/2$ state that are covered by both LRIS and UVES spectra.
We note that the \ion{Fe}{ii}\,$\lambda$2396 line is clearly detected
in the LRIS spectrum with a rest-frame equivalent width of $0.32\pm
0.06$ \AA, while the other three excited Fe$^+$ lines are only
marginally detected (we provide 2\,$\sigma$ upper limits).
%

\begin{deluxetable}{p{1.2in}cccc}
\tablecaption{Rest-frame Equivalent Widths of Fe\,II $J=7/2$ Transitions\label{EW}}
\tablewidth{0pt}
\tablehead{\colhead{} &  \multicolumn{4}{c}{EW$_{\rm rest}$(Fe$^+$,\,$J=7/2$) [\AA]\tablenotemark{a,b}} }
\startdata
$\lambda$ [\AA] & 2333.516 & 2365.552 & 2389.358 & 2396.356 \nl
$f$             & 0.06917  & 0.04950  & 0.08250  & 0.26730 \nl   
\hline
epoch 1 / LRIS & $< 0.12$ & $< 0.12$ & $< 0.12$ & $0.32\pm 0.06$ \nl
epoch 2 / UVES & $< 0.11$ & $< 0.11$ & $< 0.11$ & $< 0.11$
\enddata
\tablenotetext{a}{Systematic errors in the continuum placement are included. 
They mainly affect the LRIS spectrum, while the UVES spectrum is dominated by 
the random noise.}
\tablenotetext{b}{Upper limits are statistical, 2\,$\sigma$ non-detections.}
\end{deluxetable}
%

\begin{figure}[!]
\centering
\includegraphics[width=5in]{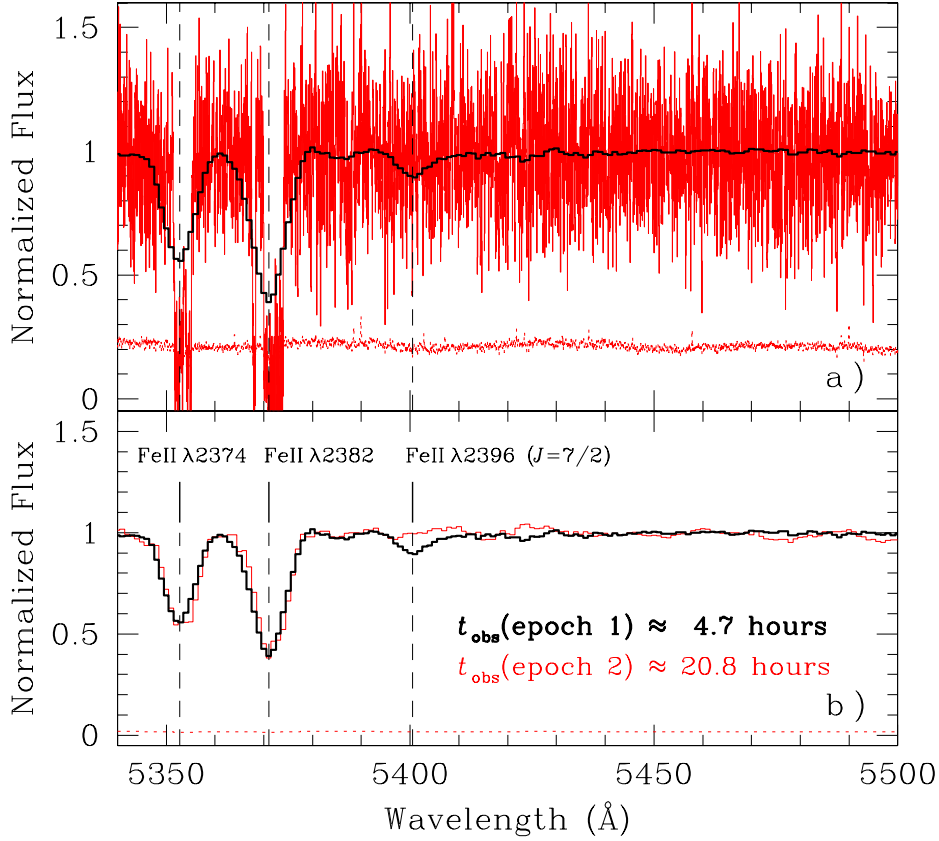}
\caption{Panel~a) Full resolution UVES spectrum of the GRB\,020813 afterglow 
with its 1\,$\sigma$ error spectrum (thin curve) acquired $\approx 20.8$ hours 
after the burst and overplotted on the low-resolution LRIS spectrum (thick 
curve) obtained only $\approx 4.7$ hours after the burst. Panel~b) Rebinned and 
smoothed 
UVES spectrum (thin curve) to the spectral resolution of the LRIS spectrum 
(thick curve) with its corresponding 1\,$\sigma$ error spectrum (warning: the 
vertical scale is different from the one used in panel~a)). 
The dashed lines correspond to the central positions of 
\ion{Fe}{ii}\,$\lambda$2374, \ion{Fe}{ii}\,$\lambda$2382, and the excited 
line \ion{Fe}{ii}\,$\lambda$2396, respectively. \ion{Fe}{ii}\,$\lambda$2396 is 
clearly detected in the epoch~1 LRIS spectrum, but is not observed in the 
epoch~2 UVES spectrum (see Table~\ref{EW} for the rest-frame equivalent width 
measurements).}
\label{UVES-LRIS}
\end{figure}
%

We searched for these Fe$^+$ excited state transitions in the epoch~2
UVES spectrum (the \ion{Si}{ii}\,$\lambda$1533 ($J=3/2$) line is not
covered by the UVES spectrum).  In contrast, the
\ion{Fe}{ii}\,$\lambda$2396 line, which is well detected in the
epoch~1 LRIS spectrum, is not present in the epoch~2 UVES spectrum.
This is illustrated in Figs.~\ref{UVES-LRIS}\,a) with the full
resolution UVES spectrum and b) with the UVES spectrum smoothed and
rebinned to the spectral resolution of the LRIS spectrum. We note that
the LRIS spectroscopic observations were carried out using a D560
dichroic.  The spectral break between the blue and red arms is located
at roughly 5600 \AA, much beyond the wavelengths where the excited
\ion{Fe}{ii}\,$\lambda$2396 features are identified.  We have also
examined the 12 individual exposures from the LRIS observations.
The Fe$^+$ absorption feature is detected in all exposures, and have
varied only very marginally over the three hours of the spectroscopic
observations.

We measured an upper limit to the rest-frame equivalent width of
\ion{Fe}{ii}\,$\lambda$2396 in the UVES spectrum over the line profile
observed in the LRIS spectrum.  A comparison between the equivalent
widths derived from the two-epoch spectra
shows that the Fe$^+$ rest-frame equivalent width in epoch~1 of 0.32
\AA\ differs from the rest-frame equivalent width in epoch~2 at the
4\,$\sigma$ significance level.  In other words, at $>99.999$\%
confidence level, the strength of the \ion{Fe}{ii}\,$\lambda$2396 line
varied between the LRIS and UVES observations.  Including a
conservative estimate of the systematic error associated with
continuum placement, the result remains at $\sim 3$\,$\sigma$
significance level (see Table~\ref{EW}).

It is, however, necessary to test the robustness of the observed
temporal variation in the excited \ion{Fe}{ii}\,$\lambda$2396 line,
because of the poor signal-to-noise ratio of the UVES data and because
of the different spectral resolutions between the two data sets.  We
therefore performed a Monte-Carlo analysis.  We first adopted the LRIS
spectrum as the fiducial model spectrum of the afterglow (given its
very high signal-to-noise ratio) and assumed that the random noise of
the UVES data is the dominant source of errors.  Next, we generated a
thousand synthetic spectra through random sampling the LRIS data
within the UVES uncertainties.  Specifically, for each pixel $i$, the
synthetic flux was obtained by drawing a random value within the
Gaussian error distribution function centered at the fiducial LRIS
flux. The width of the Gaussian distribution function was set by the
corresponding 1\,$\sigma$ error of the pixel $i$ in the UVES spectrum.

\begin{figure}[!]
\centering
\includegraphics[width=5in]{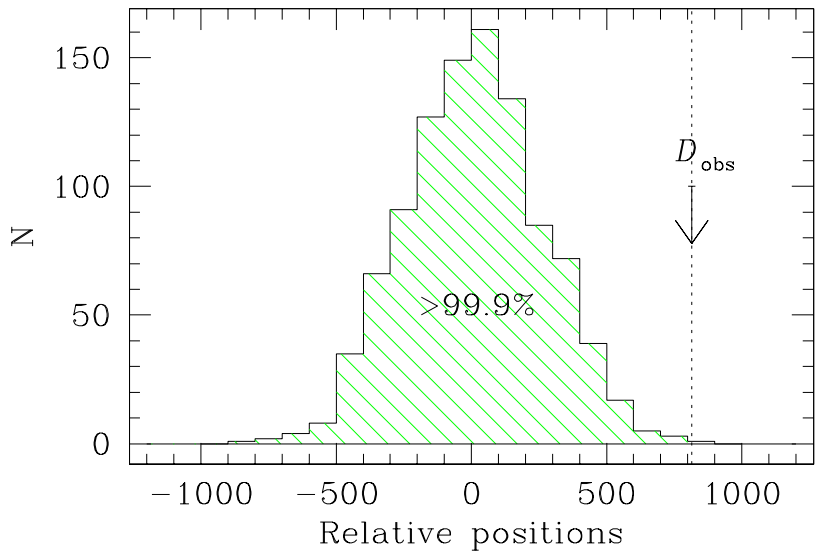}
\caption{Distribution of the relative positions between 1000 synthetic spectra
and the LRIS spectrum. The synthetic spectra were generated by random sampling 
of the LRIS data within the UVES noise (after the rebin of the UVES spectrum), 
and the relative position is defined as the sum of flux differences between two
spectra over the pixel range along the \ion{Fe}{ii}\,$\lambda$2396 ($J=7/2$) 
line profile. The dotted line corresponds to the observed relative position, 
$D_{\rm obs}$, between the rebinned UVES spectrum and the LRIS spectrum. 
We find that $> 99.9$\% of the chances $D<D_{\rm obs}$ in the distribution.}
\label{monte-carlo}
\end{figure}

We applied the Monte-Carlo analysis to the rebinned 
UVES spectrum, i.e., at the same spectral resolution.
To test whether or not the excited \ion{Fe}{ii}\,$\lambda$2396
($J=7/2$) line observed in the LRIS spectrum could have been visible
at the noise level of the UVES data, we considered ``the relative
position'' defined as the sum of flux differences between two spectra
over the pixel range along the \ion{Fe}{ii}\,$\lambda$2396 line
profile. In Fig.~\ref{monte-carlo}, we plot the distribution of
relative positions between the synthetic spectra and the LRIS
spectrum. We compare it to the observed relative position, $D_{\rm
obs}$, between the UVES spectrum and LRIS spectrum. We find that
$<0.1$\% of the trials give $D_{\rm obs}$ as a result of random noise.
This indicates that the UVES spectrum is statistically different from
the LRIS spectrum
and that the \ion{Fe}{ii}\,$\lambda$2396 line in the epoch~2 spectrum is 
significantly weaker than what is detected in epoch~1 spectrum.  The observed 
difference is at greater than 3\,$\sigma$ level of significance.

The robustness of this result can be checked by considering a
``control'' spectral interval, chosen to be outside the
\ion{Fe}{ii}\,$\lambda$2396 line profile and containing just the
continuum. The Monte-Carlo analysis over this control interval shows
that $D_{\rm obs}$ lies in the middle of the distribution of relative
positions between the synthetic spectra and the LRIS spectrum --~about
50\% of values in the distribution are greater than $D_{\rm obs}$ and
about 50\% are lower. Hence, the UVES and LRIS spectra cannot be
distinguished from each other in this control interval.
%

%

\begin{figure}[!]
\centering
\includegraphics[width=3in,angle=270]{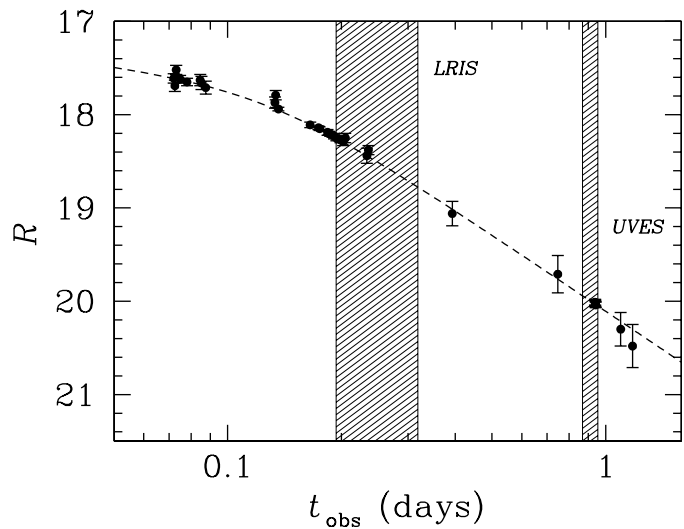}
\caption{Observed $R$-band light curve for GRB\,020813.  Different photometric 
data are collected from \citet{gladders02}, \citet{kiziloglu02}, \citet{li03}, 
and \citet{urata03}.  The dashed curve is the smooth, double power-law model 
derived by \citet{li03} that best represents the decline of the afterglow 
light.  Shaded area indicates the epochs when the spectroscopic data were 
taken.}
\label{lightcurve}
\end{figure}
%

\section{Discussion}

We have compared afterglow spectra of GRB\,020813 obtained roughly 16
hours apart.  Significant variation in the strength of the excited
Fe\,II\,$\lambda$2396 feature identified in the GRB host is detected
at $>3$\,$\sigma$ level of significance.  A natural consequence of the
excited lines being produced by UV pumping is that these excited
levels are populated according to the flux of the radiation field. As
the GRB afterglow fades, one must witness the decay of the excited
states (PCB06).  Our analysis of the afterglow spectra of GRB\,020813
presents the first supporting evidence for the UV pumping model.

For the Fe$^+$ excited levels, the half-life for spontaneous emission
is $\approx 10$ minutes, significantly smaller than the characteristic
decay time of the GRB afterglow (Fig.~\ref{lightcurve}).  Therefore,
the population of the excited states of Fe$^+$ is expected to track
the UV flux.  At $t_{\rm obs} \approx 4.7$ and $t_{\rm obs} \approx
20.8$ hours, the excitation rates of the \ion{Fe}{ii} $J=7/2$ level
due to indirect UV pumping are $\approx 25$ s$^{-1}$ and 5 s$^{-1}$,
respectively.  The excitation rates are estimated adopting the
temporal declining index $\alpha=-1.09$ from \citet{li03} and a
spectral index $\beta=-1$ from \citet{covino03}.  They are roughly
consistent with the measured equivalent width and limit from the
two-epoch observations.

Following PCB06, we find that the Fe$^+$ gas must lie beyond 40 pc
from the GRB afterglow in order to survive the intense ionizing
radiation field, but within 100 pc of the GRB afterglow such that it
will significantly populate the Fe$^+$ levels at $t_{\rm obs} \approx
4.7$ hours. In the case of GRB\,020813, the absence of
\ion{Fe}{ii}\,$\lambda$2396 at $t_{\rm obs} \approx 20.80$ hours
further constrains the Fe$^+$ gas at $> 50$ pc from the afterglow.
This is consistent with the presence of Mg$^0$ identified in the host
environment \citep{barth03,savaglio04} which implies a distance $>30$
pc (PCB06). The results presented here stress the value of obtaining
both early and late-time observations of GRB afterglows.  While
acquiring the observations using the same telescope and instrument is
preferred, we demonstrate that precise equivalent width measurements
using data from different facilities can also yield strong constraints
for the temporal evolution of the excited levels.  Finally, a
time-dependent calculation of radiative transfer and ion excitation
may help further reveal the geometry and physical conditions of the
gas.


%

\acknowledgements

The authors wish to acknowledge the ESO Science Archive Facility for
the access to the UVES/VLT spectrum of GRB\,020813.  M.D.-Z. extends
special thanks to S. Udry and F. Pont for useful discussions.
H.-W.C. thanks ESO (Garching) for their hospitality during the writing
of this manuscript.  J.X.P. acknowledges helpful discussions with
B. Mathews and R. Guhathakurta.  J.X.P., H.-W.C., and J.S.B. are
partially supported by NASA/Swift grant NNG05GF55G.
%

\end{document}